# Nitrobenzene as Additive to Improve Reproducibility and Degradation Resistance of Highly Efficient Methylammonium-Free Inverted Perovskite Solar Cells

**Apostolos Ioakeimidis [1] and Stelios. A. Choulis [1], ***

[1] Molecular Electronics and Photonics Research Unit, Department of Mechanical Engineering and Materials Science and Engineering, Cyprus University of Technology, 45 Kitiou Kyprianou Street, Limassol, 3603, Cyprus; a.ioakeimidis@cut.ac.cy

* Correspondence: stelios.choulis@cut.ac.cy

**Abstract:** We show that the addition of 1 % (v/v) nitrobenzene within the perovskite formulation can be used as a method to improve the power conversion efficiency and reliability performance of methylammonium-free (CsFA) inverted perovskite solar cells. Addition of nitrobenzene increased PCE due to defect passivation and provides smoother films resulting in PVSCs with narrower PCE distribution. Moreover, the nitrobenzene additive methylammonium-free hybrid PVSCs exhibit prolonged lifetime compare to additive free PVSCs due to enhanced air and moisture degradation resistance.

## 1. Introduction

The power conversion efficiency (PCE) of hybrid perovskite solar cells (PVSC) has exceeded 25% reaching this of silicon technology [1]. Nevertheless, the short lifetime and low reproducibility are major obstacles that prevents the commercialization of hybrid perovskite technology. Various degradation protocols are applied to predict the long lifetime of the devices such as continues illumination, heat, humidity etc[2,3]. Regarding the humidity resistance the strategies that have been developed to improve the performance include the better encapsulation, application of hydrophobic back contact buffer layers, post treatment of perovskite with functional molecules, use of additive in the perovskite solution etc [4–10]. The use of additives is a simple solution-based approach and thus a great variety of functional molecules and polymers have been investigated to improve both the efficiency and stability performance of PVSC [11–14]. In principles the various molecules used for the humidity resistance improvement composed of two main chemical groups, one that can coordinate with the perovskite and the other it can either coordinate with perovskite or it can be hydrophobic [15].

In one of the first reports regarding the use of additive for the enhancement of PVSC, Po-Wei Liang et. al. investigated the use of 1,8-diiodooctane (DIO) to improve the film quality by controlling crystallization process (nucleation and growth of crystallites) and by improving perovskite's morphology and coverage the PCE of the corresponding devices was increased by ~31%[16]. Xiong Li et. al. reported the improvement of PCE and stability of PVSC using phosphonic acid ammonium molecules. It was shown that strong hydrogen bond was formed between $-PO(OH)_2$ of the additive with the perovskite terminal groups $-NH_3^+$ resulting in a smooth and dense perovskite layer. The PCE of the corresponding PVSC incorporating butylphosphonic acid 4-ammonium chloride compare to additive free PVSCs was doubled and exhibited improved resistance to moisture degradation [17]. In another report Ignasi Burgués-Ceballos et. al.

systematically investigated the impact of various additives on the perovskite ( commercial precursor ink (I201) from Ossila Ltd) film morphology and on the performance of PVSCs concluding that the addition of small amount (1-5% v/v) of benzaldehyde can increase the PCE by 10%[18]. Since then several different molecules have been applied for additive engineering of perovskite formulations [19–27]. In one of the few reports where methyl ammonium free perovskite was engineered applying molecular additives is by Chao Shen et. al. who proposed the addition of sulfonyl fluoride-functionalized phenethylammonium in precursor to stabilize FAPbI$_3$ perovskite. It was shown that the interaction of sulfonyl group leads to improved crystallinity and passivated surface defects inducing an increased resistance to moisture invasion. As a result, the additive based FAPbI$_3$ perovskite PVs remained stable in air for more than 1000 hours, while the reference devices without additive exhibited a severe reduction in the PCE within the first 100 hours of the measurement [28].

Another critical issue for controlling the properties of the perovskite formulation is the selection of perovskite composition which also affects the stability to the various degradation environments [29]. Among them, methylammonium-free (CsFA) perovskite emerges as a potential formulation for efficient and stable PVSC since it does not incorporate the volatile methylammonium cations [29–35]. Michael Graetzel et. al. has reported that hybrid PVSCs with composition $Cs_{0.2}FA_{0.8}PbI_{2.84}Br_{0.16}$ show a stabilized PCE of unencapsulated devices in air for more than 1000 hours [36–38].

Here, we report the use of nitrobenzene additive into solution processed hybrid methylammonium-free (CsFA) perovskite as a method to improve PCE, reliability and air/humidity resistance of inverted PVSCs. The selection of the specific molecule is based on previous reported literature that the nitro group can interact with PbI$_6$ cage of the perovskite and can lead to a passivation effects, while the benzene group is a hydrophobic group that has the potential to protect from moisture ingress [39,40]. We show that the addition of nitrobenzene-based additives within the methylammonium-free (CsFA) hybrid perovskite formulation results to increased mean PCE with much narrower distribution compare to additive-free PVSC under investigation. The increased open circuit voltage (Voc) of hybrid PVSC by using 1 % (v/v) nitrobenzene additive within the formulation of methylammonium-free (CsFA) indicates the defect passivation. The presented UV-Vis measurements on hybrid PVSC precursor solutions that contains nitrobenzene suggests the interaction of perovskite's colloidal particles with nitrobenzene, while the topography of the nitrobenzene based PVSC active layer show a reduced roughness/thickness inhomogeneity as well as passivated grain boundary defects. Moreover, the nitrobenzene additive methylammonium-free (CsFA) hybrid PVSCs retain over 85% of their initials mean PCE after 1500 hours in air whereas the additive-free hybrid PVSC under investigation decline by more than 65 %. Accelerated humidity lifetime testing that performed in humidity chamber at 75% RH and 22ºC combined with photocurrent mapping measurements has further shown that nitrobenzene based methylammonium-free (CsFA) inverted PVSC are more stable due to the defect passivation and inhibition of moisture permeation effects.

## 2. Materials and Methods

*2.1. Materials*

Pre-patterned glass-ITO substrates (sheet resistance 4 ohm.sq$^{-1}$) were purchased from Psiotec Ltd., PbI$_2$ from Alfa Aesar, PC$_{60}$BM from Solenne BV and PEDOT:PSS (Al 4083) solutions from the Heraeus Clevios™. All the other chemicals used in this study were purchased from Sigma-Aldrich.

*2.2. HTLs Preparation on ITO*

For the NiOx HTLs a solution combustion process similar to Jae Woong Jung et. al. was followed[41]. In details, 1 mmol Ni(NO$_3$)$_2$.6H$_2$O, were dissolved in 1 mL 2-methoxy ethanol by stirring for about 15 min. Then, 30.65 μL Acetylacetonate were added to the solution and stirred for 1 hour in room temperature. ITO substrates were sonicated in acetone and subsequently in isopropanol for 10 min and dried by blow with nitrogen gas before use. Doctor blade technique was applied for the fabrication of the precursor films and the resulting films were dried at 100 °C for 5 min. Subsequently, the obtained films were heated at 300 °C in air for 1 h on a preheated hot plate to complete the combustion process. For PEDOT:PSS (AI 4083) the solution was used without any further treatment. The fabrication of PEDOT:PSS was performed in air where 50μL were spin-coated on ITO using static method at 4000rpm for 30 sec and then annealed at 150°C for 15 min. All films after the annealing were left to cool down at room temperature for at least 5 min and then transferred into the glovebox.

*2.3. Devices Fabrication*

The inverted PVSC under study was ITO/NiOx/ Cs$_{0.17}$FA$_{.0.83}$Pb(I$_{0.87}$Br$_{0.13}$)$_3$/PC$_{60}$BM/BCP/Cu. The perovskite solution was prepared in glove box by mixing 484.4 mgr PbI$_2$, 93.42 mgr PbBr$_2$, 186.3 mgr Formamidinium iodide (FAI) and 57.6 mgr CsI in 1 ml of 4:1 dimethylformamide (DMF): Dimethyl sulfoxide (DMSO) and steered for 15 minutes at 70ºC. This solution was split in two equal parts and in one of the two same solutions 1 % v/v of nitrobenzene was added and then both solutions were steered for 10 more minutes at 70ºC. All the required steps for the PVSC fabrication, after NiOx (PEDOT:PSS) deposition, were performed in the glovebox with >1ppm O$_2$ and >3ppm H$_2$O. The methylammonium free (CsFA) perovskite active layers were fabricated applying deposition parameters similar to Kelly Schutt et. al[31]. Specifically, the perovskite precursor solution was deposited (45 μL on 1.5 x 1.5 cm substrate) on the NiOx (PEDOT:PSS) and spin-coated for 10 sec. at 1000 rpm and then for 35 sec. at 6000 rpm. During the second step and 10 sec before the end 100 μL of chlorobenzene were drop casted on the substrate and the film changed color from bright yellow to brown within the next few seconds. Then the films were annealed at 80 ºC for 1 min on hot plate followed by 100 ºC annealing for 10 min. Next, the substrates were left for 5 min to cool down and PC$_{60}$BM (20 mg/mL in chlorobenzene) solution was dynamically spin coated on the perovskite layer at 1000 rpm for 30 s without any further annealing. The substrates were transferred in a vacuum chamber without been exposed to air and then under a base pressure of ~5*10$^{-7}$ mbar a thin film of 7 nm bathocuproine (BCP) were deposited. Subsequently, the devices were competed by thermally evaporating 100 nm of Copper (Cu) through a shadow mask resulting in an active area of 0.9 mm$^2$. The Encapsulation was applied directly after evaporation in the glove box using a glass coverslip and an Ossila E131 encapsulation epoxy resin activated by 365 nm UV irradiation.

*2.4. Characterizations*

For the UV–vis, AFM absorption measurements the perovskite films were prepared on ITO/NiOx substrates. To perform PL measurements the perovskite films were prepared (as described in device fabrication section) on quartz substrates which was treated with UV-$O_3$ for 15 min prior to deposition. Absorption measurements on both films and solutions were performed with a Schimadzu UV-2700 UV–vis spectrophotometer. The thickness of the films was measured with a Veeco Dektak 150 profilometer. The current density–voltage (J/V) and Voc-intensity were obtained using a Botest LIV Functionality Test System measured with 10 mV voltage steps and 40 ms of delay time. For illumination, a calibrated Newport Solar simulator equipped with a Xe lamp was used, providing an AM1.5G spectrum at 100 mW cm−2 as measured by a certified oriel 91150 V calibration cell. A shadow mask was attached to each device prior to measurements to accurately define the corresponding device area. Steady-state PL experiments were performed on a Fluorolog-3 Horiba Jobin Yvon spectrometer based on an iHR320 monochromator equipped with a visible photomultiplier tube (Horiba TBX-04 module).The PL was non-resonantly excited at 550 nm with the line of a 5 mW Oxxius laser diode. EQE measurements were performed by Newport System, Model 70356_70316NS . AFM images were obtained using a Nanosurf easy scan 2 controller under the tapping mode. The ageing of the devices was conducted in an environmental chamber.

## 3. Results and Discussion

*3.1. Perovskite Solar Cells*

To investigate the additive engineering of methylammonium-free (CsFA) perovskite PVs with nitrobenzene two batches (12 samples in each batch for over ten repeated runs) of PVSC with the structure glass/ITO/NiOx/Perovskite/$PC_{60}BM$/BCP/Cu were prepared. For the methylammonium free (CsFA) perovskite active layers were fabricated by applying deposition parameters similar to Kelly Schutt et. al., while for the bottom electrode NiOx (hole transporting layers) HTLs were fabricated on ITO by solution combustion process similar to Jae Woong Jung et. al. [31,41]. The $PC_{60}BM$ electron transporting layer (ETL) was spin-coated followed by thermal evaporation of bathocuproine (BCP) and Copper (Cu) to complete the inverted PVSCs top electrode. Full details are provided within the materials and methods section. For the first batch the perovskite solution was prepared without any additive, while the other batch prepared with the addition of nitrobenzene in perovskite solution, where for both batches the used perovskite composition is the methylammonium-free (CsFA) $Cs_{0.17}FA_{.0.83}Pb(I_{0.87}Br_{0.13})_3$. By applying different concentrations of nitrobenzene in the methylammonium-free (CsFA) perovskite precursor solution PCE as a function of additive concentration was investigated and the optimum amount of nitrobenzene additive was identified to be 1% v/v (Figure S1). Figure 1 (a) demonstrates the PCE distributions of the two batches of PVSC with and without 1% nitrobenzene as well as the extracted mean and standard deviations, respectively. The batch with the nitrobenzene additive exhibits an increased mean PCE value 17.09% compare to reference (15.34 %) with higher reproducibility since the respective standard deviation of the former is almost half (0.64 %) compare to the last (1.15%). Further, the best performing devices of each batch are presented in Figure 1(b) and the extracted photovoltaic (PV) parameters in Table 1. The PVSC with the nitrobenzene delivers Voc = 0.92 V, Jsc = 24.36 mA/cm$^2$ and FF = 80.3 % delivering a PCE = 18.02 %, while the reference device Voc = 0.89 V, Jsc = 23.99 mA/cm$^2$ and FF = 81.3 % delivering a PCE = 17.35 %. The integrated current from the external quantum efficiency (EQE) (Figure 1(c)) is 22.78 and 23.17 mA/cm$^2$ for the nitrobenzene and reference device, respectively, in good consistent with the solar simulator extracted values. The 1% nitrobenzene device shows an enhanced photo response for the wavelengths shorter than 500 nm compare to additive-free reference devices. It will be shown later through optical absorption measurements that nitrobenzene based methylammonium-free (CsFA) perovskite PVs exhibit an increased optical absorption at this range. The experimental results provide evidence that the addition of nitrobenzene results in better control of methylammonium-free (CsFA) perovskite active layer formation while the observed increase in the reported Voc values indicates passivation of surface defects that consistently resulted to improve device performance reliability.

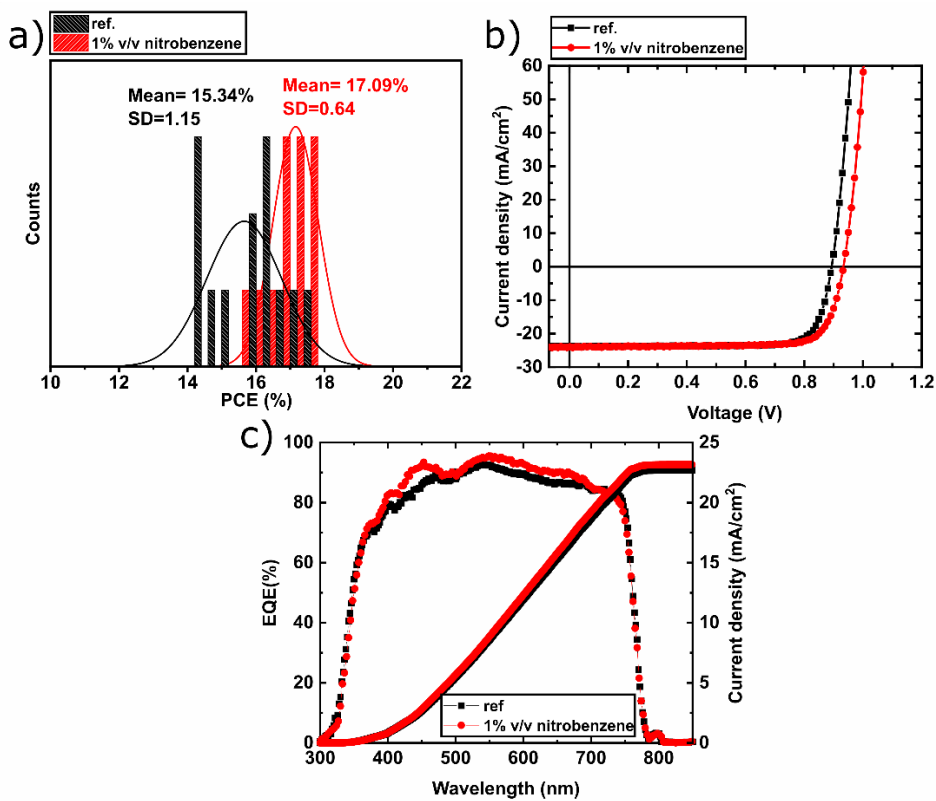

**Figure 1.** (**a**) the average PCE and standard deviation (SD) of methylammonium-free (CsFA) PVSC with and without nitrobenzene and the corresponding (b) J-V,(c) EQE and integrated current density of the best performing devices of each batch..

**Table 1.** Extracted PV parameters from the J-V curves of the best performing devices.

|  | Voc (V) | Jsc (mA/cm$^2$) | FF (%) | PCE (%) |
| --- | --- | --- | --- | --- |
| reference | 0.89 | 23.99 | 81.3 | 17.35 |
| 1 % v/v Nitrobenzene | 0.92 | 24.36 | 80.3 | 18.02 |

*3.2. Perovskite Solutions and Films Characterization*

Further measurements were performed to better clarify the effect of nitrobenzene addition into CsFA PVSCs. To probe the impact of nitrobenzene into the methylammonium-free (CsFA) perovskite formulation, UV-Vis absorption measurements were conducted on each precursor solution, with their concentration being two-thirds the concentration used for the PVSC active layer formation in order to let enough light transmitted through. Figure 2 shows the calculated Tauc-plots of the respective absorption spectra. It is revealed that the absorption band gap of the nitrobenzene based PVSC formulations are red shifted (2.61 eV) compare to reference solution (2.66 eV). As it has been shown in previous reports, the formed complex of the perovskite precursors with molecules induce a red shift of the absorption edge for the solution under study[42–45]. Similarly, the observed red shift of the nitrobenzene containing solution compare to reference is an indication of complex formation between nitrobenzene additive and the colloidal particles of the perovskite precursor solution. Additional UV-Vis measurements (Figure 2(b)) were performed on the ITO/NiOx/CsFA perovskite structure with and without nitrobenzene displaying similar spectra at the region up to ~500 nm while an increase in the absorption of the nitrobenzene containing film was observed for wavelengths shorter than 500 nm. This change is ascribed to reduced light scattering due to smoother active layer topography rather than to enhanced crystallinity, since the mean size of the grains exhibit a minor change as shown by the AFM measurements analysis that is provided below[12,46–51]. To study grain boundary passivation effects photoluminance (PL) (Figure 2 (c)) measurements were performed on methylammonium-free (CsFA) perovskite films fabricated on glass substrates. The PL intensity of nitrobenzene containing methylammonium-free (CsFA) perovskite film exhibits an increased intensity compare to reference suggesting less defect mediated charge-carrier recombination (non-radiative process) and thus less grain boundary defects[52].

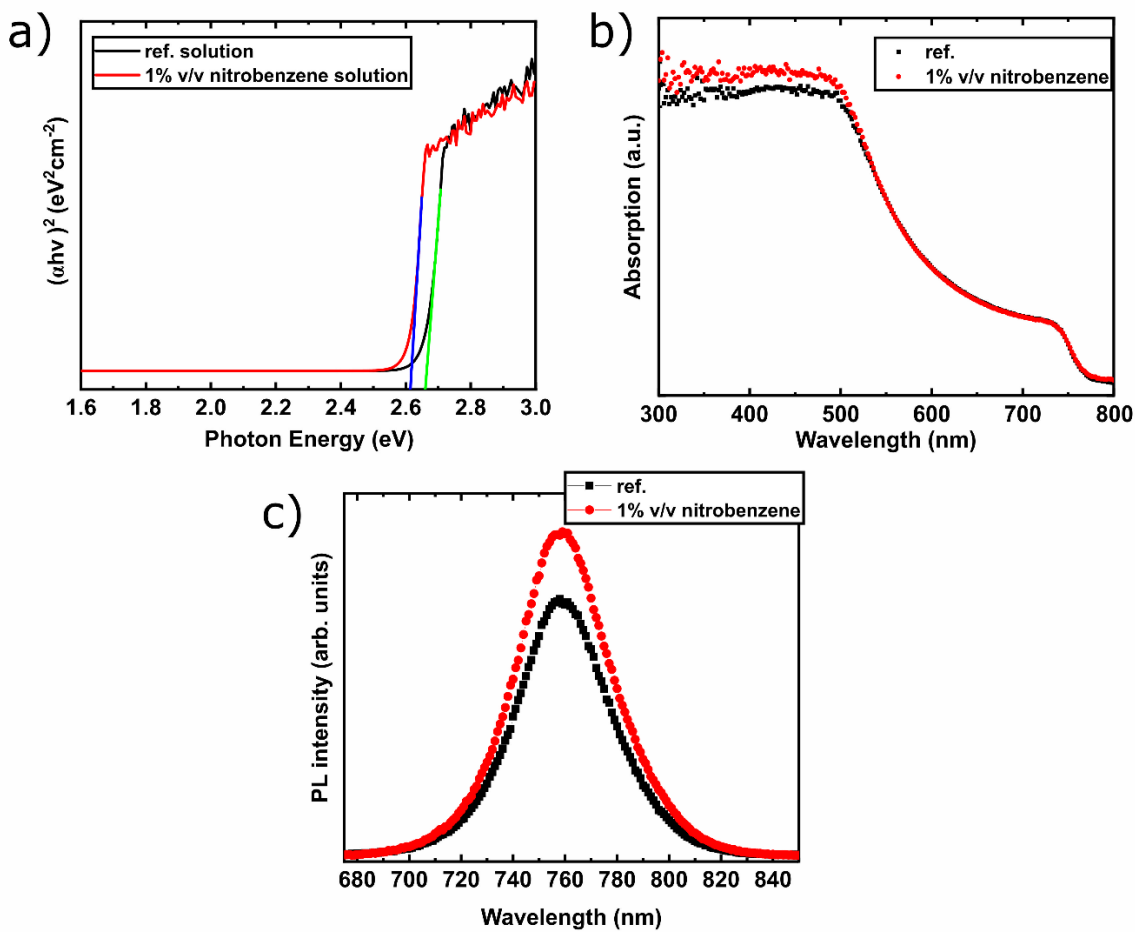

**Figure 2.** (**a**) Tauc plots of methylammonium-free (CsFA) perovskite with and without 1 % v/v Nitrobenzene additive calculated from the absorption measurements of precursor solutions and the corresponding (**b**) optical absorption and (**c**) photoluminescence (PL) of the resulting films fabricated on ITO/NiOx/CsFA and glass substrates, respectively..

To investigate the topography of methylammonium-free (CsFA) perovskite films were fabricated following the exact same processing conditions that have been applied to corresponding PVSCs on ITO/NiOx substrates with thickness ~500nm (determined by profilometry measurements). Atomic force microscopy (AFM) measurements were conducted on the corresponding active layers with the calculated roughness (root mean square) of the nitrobenzene containing active layer being reduced by ~30% compare to additive free active layer. Specifically, for 50x50 μm image (Figure 3 (a),(c)) the roughness decreases from 29.9 nm to 22.8 nm and for 10x10 μm image (Figure 3(b),(d)) from 17.7 nm to 13.5 nm. From the grains size distribution (Figure S2((a),(b)) it is calculated that the addition of nitrobenzene reduces both the mean grain size and the standard deviation to 304 nm from 329 nm and to 103 % from 118 %, respectively. Thus, the nitrobenzene containing methylammonium-free (CsFA) perovskite active layers show reduced thickness inhomogeneity and higher compactness, which can be ascribed to retarded crystal growth due to the formed complex of nitrobenzene with the precursor particles[53,54].

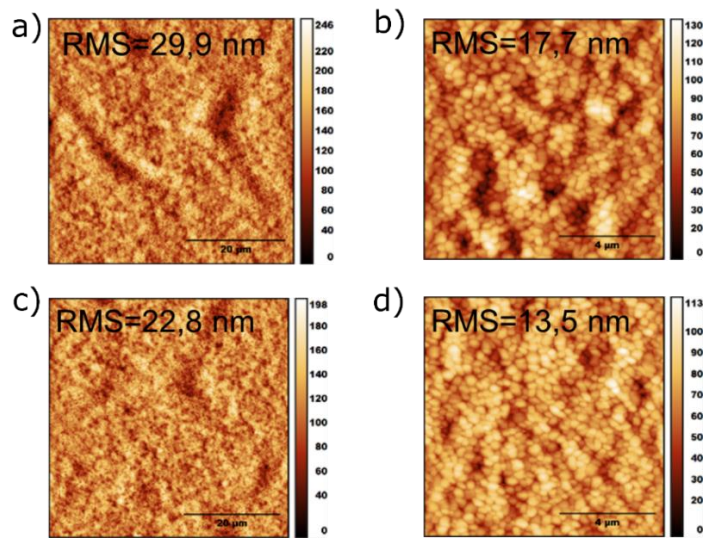

**Figure 3.** Topography pictures with sizes 50x50 μm (**a,c**) and 10x10 μm(**b,d**) obtained with AFM and the calculated roughness of the (**a,b**) reference and (c,d) nitrobenzene containing methylammonium-free (CsFA) perovskite films fabricated on ITO/NiOx substrates.

We also examined whether the improved properties of the inverted PVSC were induced by the interaction of the additive with hole transporting oxide layer (NiOx-underlayer). Since NiOx is a nanoparticulate based functional layer the high surface area/high number of surface defects could results in high reactivity with nitrobenzene which might affect the perovskite formation process[55]. Two batches of PVSC were fabricated (with and without nitrobenzene) replacing the NiOx with the PEDOT:PSS as the most common organic hole transporting layer used within inverted PVSC. Consistent to our previous reported experimental evidences using NiOx HTLs the inverted PVSCs containing nitrobenzene and PEDOT:PSS as HTL exhibit also improved PCE reproducibility with a standard deviation of 0.69 % compare to additive-free PVSCs (1.61 %) as shown in Figure S3. The above experimental results provide further indication that the origin of the improved performance of the PVSCs is due to interaction of the 1 % (v/v) nitrobenzene additive with methylammonium-free (CsFA) perovskite precursors rather than nanoparticulate metal-oxide based underlayer effects.

## 3.3. Lifetime Testing of Perovskite Solar Cells

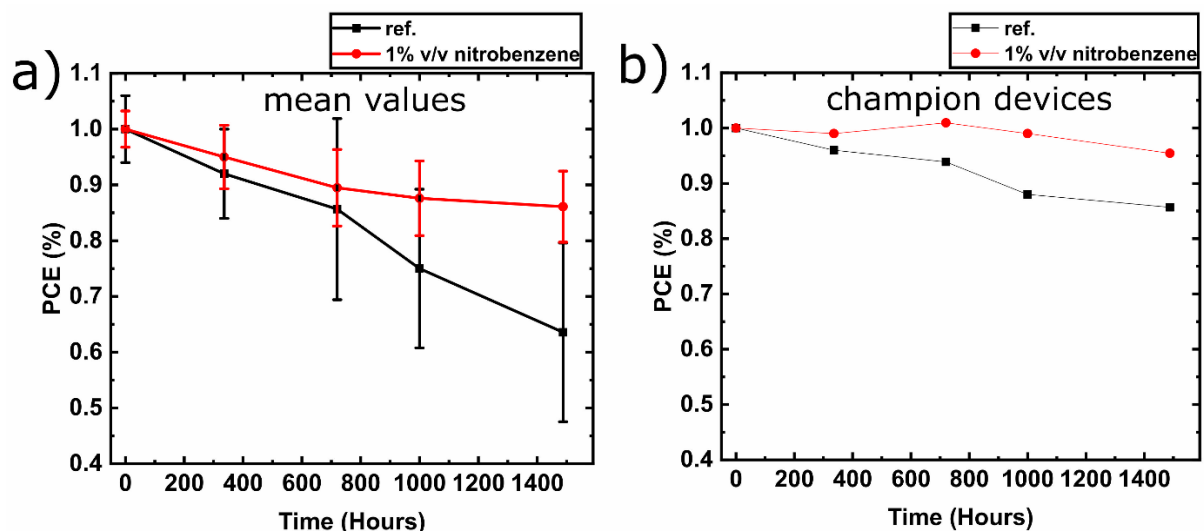

**Figure 4.** (**a**) mean PCEs and standard deviations graph of air stability measurements for the encapsulated methylammonium-free (CsFA) PVSC with and without nitrobenzene and (**b**) the corresponding champion devices.

Air stability measurements were performed on methylammonium-free (CsFA) PVSC with and without nitrobenzene, where all devices were encapsulated in inert atmosphere ($N_2$) before the exposure to air. Figure 4(a) presents the mean PCE and the standard deviation measurements throughout ageing in ambient conditions. First, it can be observed that the mean PCE of the nitrobenzene containing PVSCs retain around 85% of the initial PCEs after 1500 hour in contrast to the reference PVSCs that decline to approximately 65%. Moreover, the PCE dispersion of reference PVSCs widens significantly during the ageing test compare to nitrobenzene PVSCs. Regarding the champion devices (Figure 4(b)), the nitrobenzene containing inverted PVSCs exhibits excellent performance retaining the 95% of the initial PCE after 1500 hours in air. The results show that the addition of nitrobenzene into methylammonium-free (CsFA) perovskite improves the air stability of the corresponding encapsulated inverted PVSCs.

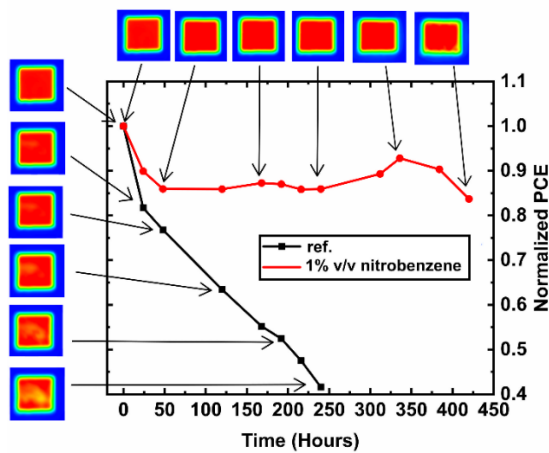

**Figure 5.** Stability measurements for the encapsulated methylammonium-free (CsFA) perovskite devices with and without nitrobenzene at 75% RH and 22 °C under dark and the respective photocurrent map of the corresponding devices.

Further to the above air stability measurements accelerated humidity lifetime testing was also investigated (75 RH% and 22 °C). During this ageing process, device lifetime performance was combined with corresponding photocurrent mapping (PCM) measurements. Figure 5 illustrates the PCE and the corresponding PCM images of the additive-free and nitrobenzene based PVSCs while the additional normalized PV efficiency parameters Jsc, Voc and FF as a function of lifetime are presented in Figure S4. The red color of the PCM represent the areas of high photogenerated current while the yellow to blue color the areas of lower photocurrent. Like the above presented air-stability measurements, the nitrobenzene containing device shows an increased humidity ageing resistance retaining the 85 % of its initial PCE for over 400 hours. Accordingly, from PCM it can be seen that the generated photocurrent shows a marginal decrease after 350 hours, mostly at the edges of the device area. On the other hand, the additive-free device shows significant decrease of its initial PCE with the first hours. The yellow spots (reduced photocurrent) at the PCM for the additive-free PVSCs observed within the first 50 hours, followed by a very aggressive photo-current degradation within the next 250 hours. The photocurrent mapping observations show good agreement with the recorded normalize Jsc values which also show an abrupt decrease during accelerated humidity testing at similar time-scales (Figure S4(b)) whereas the normalized Voc and FF (Figure S4(a),(c)) show a relative small variance within the presented lifetime performance compare to the initial values. The cause of this degradation can be ascribed to the interaction of perovskite with the $H_2O$ through the formation of the monohydrate perovskite and then the dihydrate perovskite which finally leads to its decomposition[56]. However, it remains unclear to us whether the degradation occurs at the interface of perovskite with carrier transporting layers or at the bulk perovskite (or both)[57,58]. It has been reported for other perovskite formulations (e.g. $CH_3NH_3PbI_3$, $FAPbI_3$) that the degradation initiates at the grain boundaries and propagates to the interior[58,59]. Thus, since in our report the stoichiometry of perovskite (same perovskite solution is used) and the grain sizes (shown with AFM measurements) are the same for the additive free and nitrobenzene additive based methylammonium-free (CsFA) perovskite, we infer that the origin of the enhanced degradation resistance of the nitrobenzene containing CsFA based PVSCs is due to the passivation of perovskite defects, through the reaction of nitrobenzene with the grain boundaries, as well as due to the inhibition of moisture permeation in the perovskite attributed to the hydrophobic benzene ring[13,58–63]. At this initial stage we have shown that the additive of nitrobenzene can improve the humidity life-time performance of PVSCs under 75% RH and 22°C conditions. Further ageing tests according to ISOS protocols [3] will be performed in future work.

## 4. Conclusion

In conclusion, the performance of methylammonium-free (CsFA) hybrid PVSCs that incorporate nitrobenzene additive is investigated. We have demonstrated that inverted methylammonium-free (CsFA) PVSCs using 1% v/v nitrobenzene additive provides increased mean PCE from 15.34% to 17.09%, with much narrower PCE standard deviation distribution (reduced from 1.15 % to 0.64 %) compared to corresponding additive-free PVSCs. The improved performance is attributed to the interaction of perovskite's colloidal particles with nitrobenzene as well as passivation of grain boundary defects. Importantly, the reported stability of the corresponding encapsulated air-exposed PVSCs under investigation is improved retaining 85% of the initial PCEs after 1500 hours compare to the additive-free devices which decline to approximately 65% at the same air exposure time scales. Additional accelerated humidity lifetime testing (75% RH and 22°C) show that the nitrobenzene 1% v/v containing methylammonium-free (CsFA) inverted hybrid PVSCs exhibit enhanced humidity lifetime performance , retaining the 85% of the initial PCE after more than 400 hours compare to additive-free PVSCs that decline within the first 50 hours. Although the presented lifetime measurements do not directly correspond to the ISOS based lifetime-protocols [3], the presented humidity based accelerated lifetime studies (75% RH and 22°C) combined with the photocurrent mapping measurements have shown that incorporation of nitrobenzene additive within the formulation of methylammonium-free (CsFA) hybrid inverted PVSCs can be used as a method to improve methylammonium-free (CsFA) hybrid PVSCs lifetime performance due to defect passivation and inhibition of moisture permeation.

**Acknowledgements:** This project received funding from the European Research Council (ERC) under the European Union's Horizon 2020 research and innovation program (Grant Agreement No. 647311).

# Supplementary Information

# Nitrobenzene as additive to improve reproducibility and degradation resistance of highly efficient methylammonium-free inverted perovskite solar cells

*Apostolos Ioakeimidis[1] and Stelios. A. Choulis[1*]*

[1]Molecular Electronics and Photonics Research Unit, Department of Mechanical Engineering and Materials Science and Engineering, Cyprus University of Technology, 45 Kitiou Kyprianou Street, Limassol, 3603, Cyprus

* Corespondence: stelios.choulis@cut.ac.cy

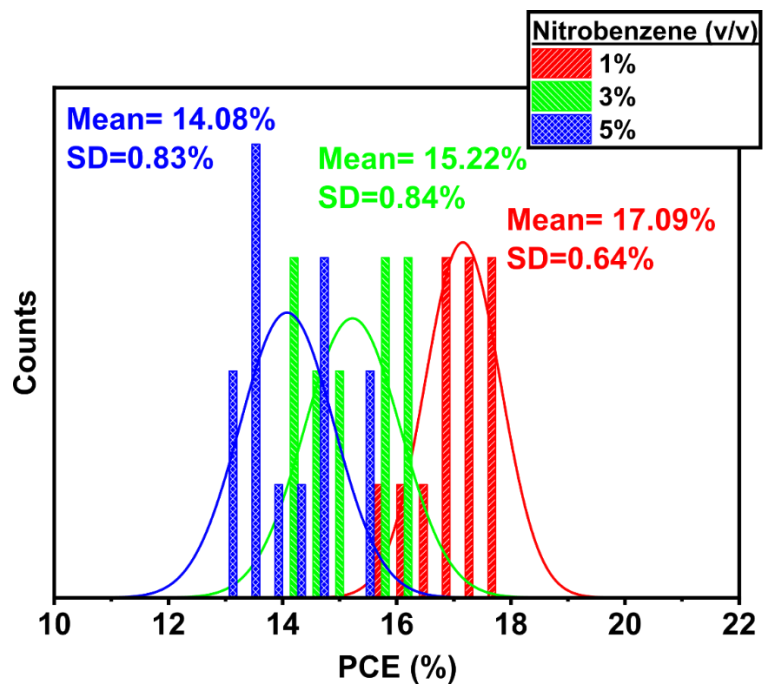

*Figure S1 mean PCE and standard deviation (SD) of PVSCs with different concentrations of nitrobenzene.*

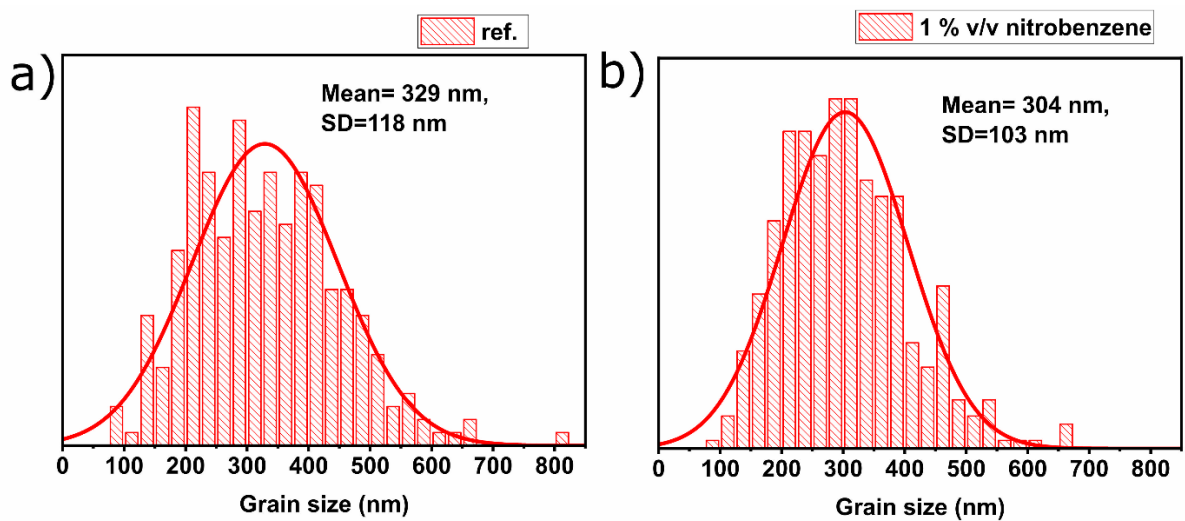

*Figure S2 Grain size distribution of pristine and nitrobenzene containing perovskite films fabricated on top of ITO/NiOx substrates*

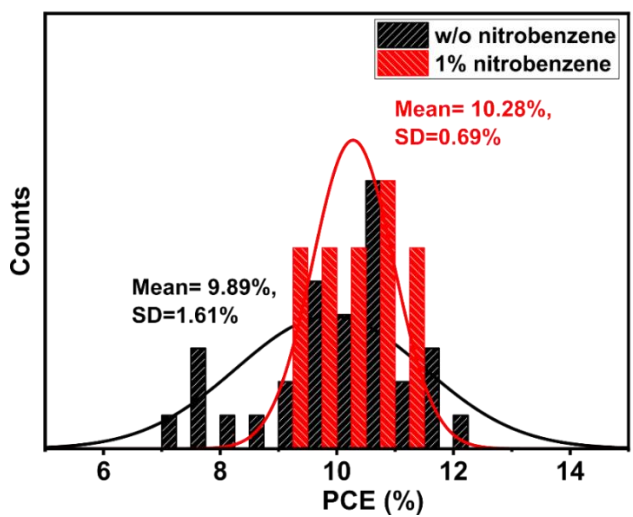

*Figure S3 mean PCE and standard deviation (SD) of PVSC with and without nitrobenzene fabricated on ITO/PEDOT:PSS substrate.*

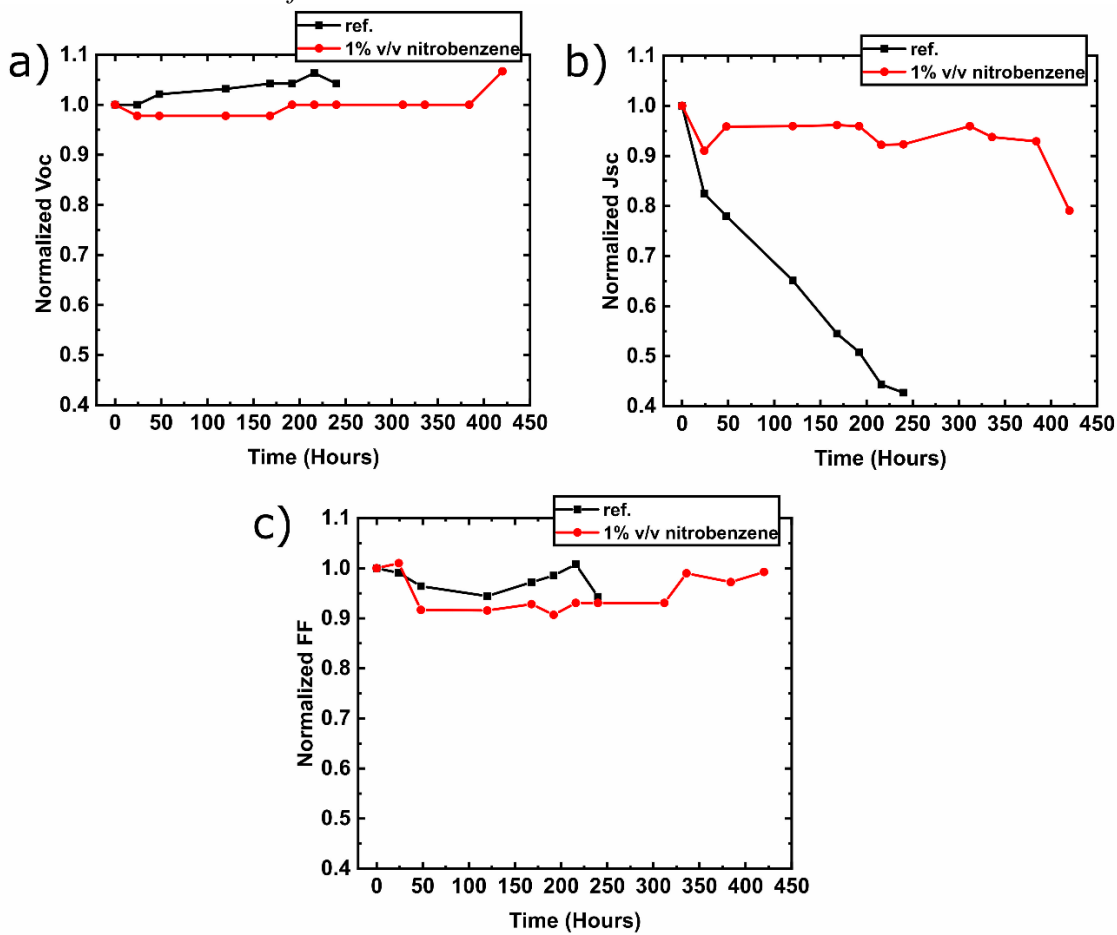

*Figure S4 The normalized PV parameters current density (Jsc), Open circuit voltage (Voc) and fill factor (FF) of the corresponding PVSC with and without 1% nitrobenzene during accelerated lifetime testing*